\documentstyle{article}
\begin{document}

\centerline{\bf Legitimacy of wave-function expansion}
\vskip 0.2in
\centerline{Dept. of Physics, Beijing University of Aeronautics}
\centerline{and Astronautics, Beijing 100083, PRC}
\centerline{C.Y. Chen, Email: cychen@public2.east.net.cn}

\vskip 0.2in
\noindent {\bf Abstract:} 
In this letter we investigate the common procedure in which any wave function is expanded into a series of eigenfunctions. It is shown that as far as dynamical systems are concerned the expanding procedure involves various mathematical and physical difficulties. With or without introducing phase factors, such expansions do not represent dynamical wave functions.


\vskip 0.2in

A common concept, which is originally due to Dirac and Neumann and has been accepted as one of the essential principles of quantum mechanics, is that any wave function can be represented by a vector in the linear infinite-dimensional vector space, the Hilbert space, whose basis is formed by a complete set of orthogonal eigenfunctions[1-4]; namely a wave function can be expanded into 
\begin{equation}\label{expansion1} \sum\limits_n C_n(t) \Psi_n({\bf r}),\end{equation} 
where $n$ stands for a set of quantum numbers labeling the eigenfunction and $C_n(t)$ a pure time-dependent complex number. The expanding procedure is also called the principle of superposition in some textbooks[4]. Adopting the expanding procedure, we can write a dynamical, nonstationary, wave function in the form (with no spin regarded)
\begin{equation} \label{expansion2}\sum\limits_n C_n(t) \exp\left(-i\frac {\varepsilon_n}{\hbar} t\right) \Psi_n({\bf r}),\end{equation}
where $\varepsilon_n$ is the eigenenergy of the system before a certain initial time $t_0$ and $\Psi_n({\bf r})$ the eigenfunction in terms of the Hamiltonian $H(t_0)$. By substituting (\ref{expansion2}) into the time-dependent Schr\"odinger equation and making use of the orthogonality of the eigenfunctions, we arrive at a set of coupled ordinary-differential equations 
\begin{equation}\label{ordi} i\hbar \frac {dC_n}{dt}= \sum_l C_l(t)(H_1)_{nl} \exp
(i\omega_{nl}t) ,\end{equation}  
where $H_1= H(t)-H(t_0)$ represents the variation of the Hamiltonian (not necessarily small). 

It is almost unanimously believed that (\ref{expansion1}) or (\ref{expansion2}) formally represents the real wave function, and that the set of ordinary-differential equations (\ref{ordi}) is equivalent to the corresponding Schr\"odinger equation. Furthermore, the value of $|C_n|^2$ is, in Dirac's perturbation theory, interpreted as a transition probability from one quantum state to another. 

While the expanding procedure outlined above seems stringent and flawless, many difficulties that can be traced back to it have constantly bothered people in the community. Several decades ago it is noticed that the resultant formula derived from the expanding procedure in Dirac's perturbation theory is not gauge-invariant[5]. Later, some in the community found that coefficients of such expansion do not represent transition probabilities[6]. To cope with these difficulties, people kept on coming up with remedies. The preferential gauge, in which the vector potential vanishes whenever the electromagnetic field becomes zero, was proposed after the debate on the gauge aspect of Dirac's perturbation theory[7]. Furthermore, it was suggested that a certain phase factor should be introduced to the wave function before using the standard expanding procedure[8]. 

In our view, those remedies have obscured, to some extent, the essence of the problem[9]. Actually, we can easily  see the situation is unsatisfactory with or without these remedies: if we adopt the expanding procedure as it is, we take the risk to confront with other physical principles such as the gauge invariance; if we believe that introducing extra additives can make the procedure work as it should, the expanding procedure becomes more a recipe than a basic physical principle. 

The objective of this letter is to directly investigate the validity of the expanding procedure. It will be illustrated that as far as dynamical systems are concerned the expansion form expressed by (\ref{expansion1}) or (\ref{expansion2}) does not represent a true dynamical wave function. If we forcefully use such expansions, abnormal things will emerge. Furthermore, it will be shown that introducing extra phase factors does not really help. 

It is very obvious that conclusions concerning this subject are crucially important since many theories and many discussions in the literature are explicitly or implicitly related to the expansion principle discussed here.

Before starting with our main discussion, it is appropriate to recall how the expanding procedure works for a stationary system. For such system, we have a time-independent Hamiltonian $H_0$ and eigenfunctions $\Psi_n({\bf r})$, and any wave function of the system can indeed be expressed by
\begin{equation}\label{sum0} \Psi(t,{\bf r}) = \sum\limits_n C_n 
\exp\left( -i\frac {\varepsilon_n}\hbar t\right) \Psi_n({\bf r}),\end{equation} 
where $C_n$'s are pure numbers independent of time and space. Note that the normalization condition 
\begin{equation}\label{summ} \sum\limits_n |C_n|^2 =1\end{equation}
holds for (\ref{sum0}) and it ensures the unity of the total probability and the convergence of the expansion.

We now try to determine whether or not a wave function of nonstationary system can be similarly expanded. We will first investigate mathematical arguments concerning this issue.

Consider a quantum system that has a stationary Hamiltonian $H_i$ before the initial time $t_i$ and another stationary Hamiltonian $H_f$ after the final time $t_f$. During the time period $t_i<t<t_f$, the system
involves a dynamical change. In using an expansion of the form (\ref{expansion1}) to describe the wave function of the system, we need to select a set of eigenfunctions we wish to use. We may choose the set of eigenfunctions of $H_i$ which we denote by $\{\Psi_{n_i}\}$, or we may equally choose the set of eigenfunctions of $H_f$ which we denote by $\{\Psi_{n_f}\}$. Referring to the discussion on the stationary system, we know that the wave function can be expressed as, before $t_i$, 
\begin{equation} \sum\limits_{n_i} C_{n_i} \exp\left( -\frac {\varepsilon_{n_i}}\hbar t\right) \Psi_{n_i}\end{equation}
and after $t_f$ 
\begin{equation} \sum\limits_{n_f} C_{n_f} \exp\left(-\frac {\varepsilon_{n_f}}\hbar t\right) \Psi_{n_f}.\end{equation}
In harmony with the expansion (\ref{expansion2}), we may simply use the eigenfunctions $\{\Psi_{n_i}\}$ to expand the wave function throughout the dynamical process. To ensure all the things go smoothly, we need to verify that any one of $\{\Psi_{n_f}\}$ can be expressed by $\{\Psi_{n_i}\}$. If this is possible, we will say the two sets of eigenfunctions, or the two states, are compatible. 

It is found that for some dynamical systems the final state is incompatible with its initial state. As an explicit example, we take a look at a particle that is trapped by a magnetic field initially and then moves in a free space since the magnetic field disappears. An  eigenfunction for the particle in a uniform magnetic field is[4]
\begin{equation}\label{mag} \Psi_{n,l,k_z}(\rho,\varphi, z)=\frac {1} 
{\sqrt{2\pi}}R_{n, l}(\rho)e^{il\varphi}e^{ik_z z} \end{equation}
where 
$$R_{n,l}(\rho)=\frac{ T(n,l) \rho^{|l|}} {a^{1+|l|}}
e^{(- \frac{\rho^2}{4a^2})} F(-n,|l|+1, \rho^2/2a^2) $$
with
$$T(n,l)=\frac 1{|l|!}\left[ \frac {(|l|+n)!}{2^{|l|}
n !} \right]^{\frac 12}$$
and $F$ is the confluent hypergeometric function 
\begin{equation} F(\alpha,\gamma,u)= 1+\frac\alpha\gamma \frac u{1!}+
\frac{\alpha(\alpha+1)}{\gamma
(\gamma+1)}\frac{u^2}{2!}+\cdots\cdots.\end{equation}
On the other hand, an eigenfunction for a free particle motion is[3]
\begin{equation}\label{plane} (8\pi^3)^{-1/2}\exp(i{\bf k}\cdot{\bf r}).\end{equation}
If we particularly choose a plane wave as
\begin{equation} \Psi_{k^\prime}({\bf r}) =C^\prime \exp (ik^\prime_z z), \end{equation}
and expand it into  
\begin{equation}\label{pse} \Psi_{k^\prime}({\bf r}) = \sum\limits_{n,l,k_z}
C_{n,l,k_z} \Psi_{n,l,k_z}. \end{equation}
where $\Psi_{n,l,k_z}$ is an eigenfunction of the type (\ref{mag}). The coefficient $C_{n,l,k_z}$ has three parameters. For the purpose of this letter, we set $l=0$ and will be concerned with the convergence of the series in terms of the index $n$ only. Along this line, we obtain 
\begin{equation}  C_{n,0,k_z} \propto \frac {\delta(k_z-k^\prime_z)}a
\int e^{-\frac{\rho^2}{4a^2}} F(-n,1,\rho^2/2a^2_h)\rho d\rho.  \end{equation}
It happens that the expression above can be evaluated analytically and the result is, with the delta-function disregarded,
\begin{equation}  C_{n+1,0,k_z}=C_{n,0,k_z}.  \end{equation}
which means that the series of the form (\ref{pse}) is not convergent in terms of the non-negative integer $n$.

Actually, the difficulty can be seen just after we write down the two sets of eigenfunctions, (\ref{plane}) and (\ref{mag}). In the $x-y$ plane,  eigenfunctions (\ref{plane}) take finite values at the infinity while eigenfunctions (\ref{mag}) tend to zero at the infinity. It is obvious that these two kinds of eigenfunctions cannot express each other. Also note that the example presented here possesses a certain degree of generality. If we deal with a situation in which a magnetic-type field is applied to an atomlike system (having  discrete and continuous spectrum), we will encounter similar difficulties.

If one thinks that the argument presented above is not general enough, look at the following argument in which we show that the normalization of wave function is broken down with the expanding procedure employed. 

We first write down the formal integration of differential equations (\ref{ordi}). Let the disturbance time interval be $t_N-t_0$, where $t_0$ and $t_N$ denote the initial time and the final time at which the perturbation turns on and off respectively. Slice the time interval into $N$ small equal increments
\begin{equation} t_0,\quad t_1,\quad t_2,\; \cdots\cdots, t_N, \end{equation}
where $t_{i+1}-t_i=(t_N-t_0)/N= \Delta t$. The coefficients $C_k$ can be evaluated by 
\begin{equation}\label{int1} \begin{array}{l}
C_k(t_1)\approx C_k(t_0)-(i/ \hbar)
\sum\limits_{k^\prime}C_{k^\prime}(t_0)(H_1)_{kk^\prime} e^{i
\omega_{kk^\prime}t_0}\Delta t \\
\cdots\cdots\\
C_k(t_{i+1})\approx C_k(t_i)- (i/\hbar)
\sum\limits_{k^\prime}C_{k^\prime}(t_i)(H_1)_{kk^\prime}e^{i
\omega_{kk^\prime}t_i}\Delta t \\
\cdots\cdots. \end{array}\end{equation}
According to Euler and Cauchy[10], if we slice the entire time into a sufficiently large number of intervals or if we are just concerned with the system's behavior within a sufficiently short time, the integral expression (\ref{int1}) can be accepted as an accurate one. 

Suppose that the system is initially in one eigenstate, namely we assume
\begin{equation}\label{cc}  C_s(t_0)=1,\quad  C_k(t_0)=0\; (k\not= s),\end{equation}
then we find that at the later time $t_1$  
\begin{equation}\label{ct1} C_s(t_1)=1- \frac i\hbar (H_1)_{ss} \Delta t.\end{equation}
Since $H_1= H(t)-H(t_0)$ is a self-adjoint operator, $(H_1)_{ss}$ is a real number. (For a uniform electric field, we can assume $H_1\propto x$ and $x_{ss}$ is just the average $x$-position in the initial state.) Then, (\ref{ct1}) tells us that
\begin{equation}\label{ct2}  |C_s(t_1)|^2\ge 1.\end{equation}
For other coefficients, we have 
\begin{equation}\label{ck2} \sum\limits_{k\not=s}|C_k(t_1)|^2\ge 0.\end{equation}
Eqs. (\ref{ct2}) and (\ref{ck2}) lead us to
\begin{equation} \sum\limits_k |C_k(t_1)|^2\ge 1.\end{equation}
The equality sign holds only when all the matrix elements of $H_1$ are zero and the system gets no disturbed. In that case, we investigate the value of the same expression at the time $t_2$. In such a way, we will surely find  
\begin{equation} \sum\limits_k |C_k(t_i)|^2> 1.\end{equation}
This clearly shows that the expansion of the form (\ref{expansion2}) or (\ref{expansion1}) is not the correct formal solution of the time-dependent Schr\"odinger equation. (In this letter, we take it for granted that the Schr\"odinger equation preserves the unity of the total probability.) 

One interesting question arises immediately: what does a dynamical wave function look like? The discussion above has suggested that such wave function must involve coupled time-and-space dependence, which cannot be handled with the variable-separation technique. (Finally, incompatible functions may also get involved.) Functions that can be expressed by (\ref{expansion1}) or (\ref{expansion2}) constitutes a special type of functions; but, paradoxically, none of them represents a dynamical wave function. 

We now turn to physical arguments of the issue and in passing we will study effects of introducing phase factors.

Before doing that it is enlightening to recall the well-known concept in quantum mechanics that the gauge fields, Hamiltonian of system, phase factor of wave function and operators representing observables are not uniquely defined and they covary in accordance to ($m=Q=c=1$ in this letter)
\begin{equation} \label{set}
\left\{\begin{array}{l}
({\bf A},\Phi)\rightarrow ({\bf A}^\prime={\bf A}+\nabla f,\; \Phi^\prime=\Phi-\partial_t f)\\
H({\bf A},\Phi)\rightarrow H({\bf A}^\prime, \Phi^\prime)\\
\Psi(t,{\bf r})\rightarrow e^{if(t,{\bf r})}\Psi(t,{\bf r})\\
L({\bf p}-{\bf A},{\bf r})\rightarrow L({\bf p}-{\bf A}^\prime,{\bf r}),     \end{array} \right. \end{equation}
where $f(t,{\bf r})$ is an arbitrary differentiable function of time and space. If and only if the covariance relationship given above is observed the physical outcomes of the formalism will be strictly gauge-invariant.

We can immediately see that our expansion principle will suffer in this respect. The third equation of (\ref{set}) illustrates that even if the wave function of interest could be expressed by (\ref{expansion1}) or (\ref{expansion2}) in a special gauge, the wave function in a general gauge would involve an extra nontrivial phase factor $e^{if(t,{\bf r})}$. Such nontrivial phase factors pose a serious difficulty for the standard expanding procedure expressed by (\ref{expansion1}), (\ref{expansion2}) as well as the differential-equation set (\ref{ordi}). This partly explains why the time-dependent perturbation theory encounters gauge difficulties.

At this point, it is in order to point out that the standard expansion procedure is in conflict with the covariance relationship expressed by (\ref{set}). Consider a quantum particle that is bounded in a mechanical well. Its state before $t=0$ is assumed to be
\begin{equation} \Psi(t<0)= \exp\left(i\frac{\varepsilon_s}\hbar t\right)\Psi_s({\bf r}).\end{equation}
At and after $t=0$, an electromagnetic disturbance is applied to the system. If we admit the standard expansion principle, the wave function after $t=0$ must be in the form
\begin{equation}\label{psiex} \Psi(t\ge 0) = \sum\limits_l C_l(t)\exp\left(i\frac{\varepsilon_l}\hbar t\right)\Psi_l({\bf r}), \end{equation} 
where
\begin{equation} C_s(0)=1, \quad C_l(0)=0\; ({\rm for}\; l\not=s).\end{equation} 
We note that the quantum system has its own inertia (the heavier the particle the larger the inertia) so that all coefficients in the expansion, as well as physical observables associated with the system, must vary tardily in comparison with the variation of the disturbance field. The velocity operator reads according to the standard definition 
\begin{equation} {\bf v}={\bf p}-{\bf A}= -i\hbar \nabla - {\bf A}.\end{equation}
where ${\bf A}$ is the vector potential chosen to represent the electromagnetic disturbance. Before $t=0$, the average velocity of the particle is
\begin{equation} \overline{\bf v}= \langle \Psi_s|-i\hbar\nabla|\Psi_s\rangle.\end{equation}
At the instant $t=0+\epsilon$, where $\epsilon$ is a truly small time interval, the average velocity becomes 
\begin{equation} \overline{\bf v}= \langle \Psi_s|-i\hbar\nabla|\Psi_s\rangle -
\langle \Psi_s|{\bf A} |\Psi_s\rangle.\end{equation}
The trouble with the expansion (\ref{psiex}) is now obvious: not only that the average value of the velocity is gauge-dependent but also that its value varies as promptly as the vector potential. Actually, all observables, including the transition probability, suffer from the same difficulty with the use of the expansion (\ref{psiex}). 

There remains one final question concerning whether or not the expression  
\begin{equation}\label{final1} e^{if(t,{\bf r})} \sum C_n(t) \Psi_n({\bf r}) \end{equation}
represents a true dynamical wave function. Our answer to it is a negative one. If this expression represented a dynamical wave function, we could find a special gauge under which 
\begin{equation}\label{final2}  \sum C_n(t) \Psi_n({\bf r}) \end{equation}
would represent a dynamical wave function. According to the arguments presented in this letter, particularly those independent of gauge, this is not at all possible.

In conclusion, it has been shown that the common expanding procedure, or the so-called principle of superposition, holds only for stationary systems. However, we notice that in the real life many quantum theories concerning nonstationary dynamical processes have assumed the general validity of it[11].

Discussion with Professors R. G. Littlejohn and Dongsheng Guo is gratefully acknowledged. This work is partly supported by the fund provided by Education Ministry, PRC.

\end{document}